\documentclass{appolb}
\usepackage{graphicx}
\usepackage{epsfig}
\begin{document}
\title{$\beta$-delayed $\gamma$-proton decay in $^{56}$Zn: analysis of the charged-particle spectrum
\thanks{Presented at the Zakopane Conference on Nuclear Physics "Extremes of the Nuclear Landscape", August 31 - September 7, 2014, Zakopane, Poland.} }
\author{ S.~E.~A.~Orrigo$^{a}$, B.~Rubio$^{a}$, Y.~Fujita$^{b}$, B.~Blank$^{c}$, W.~Gelletly$^{d}$, J.~Agramunt$^{a}$, A.~Algora$^{a,e}$, P.~Ascher$^{c}$, B.~Bilgier$^{f}$, L.~C{\'a}ceres$^{g}$, R.~B.~Cakirli$^{f}$, H.~Fujita$^{h}$, E.~Ganio{\u{g}}lu$^{f}$, M.~Gerbaux$^{c}$, J.~Giovinazzo$^{c}$, S.~Gr{\'e}vy$^{c}$, O.~Kamalou$^{g}$, H.~C.~Kozer$^{f}$, L.~Kucuk$^{f}$, T.~Kurtukian-Nieto$^{c}$, F.~Molina$^{a,i}$, L.~Popescu$^{l}$, A.~M.~Rogers$^{m}$, G.~Susoy$^{f}$, C.~Stodel$^{g}$, T.~Suzuki$^{h}$, A.~Tamii$^{h}$, J.~C.~Thomas$^{g}$
\address{ 
$^a$Instituto de F{\'i}sica Corpuscular, CSIC-Universidad de Valencia, E-46071 Valencia, Spain \\
$^b$Department of Physics, Osaka University, Toyonaka, Osaka 560-0043, Japan \\
$^c$Centre d'Etudes Nucl{\'e}aires de Bordeaux Gradignan, CNRS/IN2P3 - Universit{\'e} Bordeaux 1, 33175 Gradignan Cedex, France \\
$^d$Department of Physics, University of Surrey, Guildford GU2 7XH, Surrey, UK \\
$^e$Inst. of Nuclear Research of the Hung. Acad. of Sciences, Debrecen, H-4026, Hungary \\
$^f$Department of Physics, Istanbul University, Istanbul, 34134, Turkey \\
$^g$Grand Acc{\'e}l{\'e}rateur National d'Ions Lourds, BP 55027, F-1407 Caen, France \\
$^h$Research Center for Nuclear Physics, Osaka University, Ibaraki, Osaka 567-0047, Japan \\
$^i$Comisi{\'o}n Chilena de Energ{\'i}a Nuclear, Casilla 188-D, Santiago, Chile \\
$^l$SCK.CEN, Boeretang 200, 2400 Mol, Belgium \\
$^m$Physics Division, Argonne National Laboratory, Argonne, Illinois 60439, USA \\
}}
\maketitle

\begin{abstract}
A study of the $\beta$ decay of the proton-rich $T_{z}$ = -2 nucleus $^{56}$Zn has been reported in a recent publication. A rare and exotic decay mode, $\beta$-delayed $\gamma$-proton decay, has been observed there for the first time in the $fp$ shell. Here we expand on some of the details of the data analysis, focussing on the charged particle spectrum.
\end{abstract}


\section{Introduction}
The $\beta$ decay of the $T_{z}$ = -2 nucleus $^{56}$Zn has been studied recently \cite{Orrigo2014}. Among the interesting results, a rare and exotic decay mode has been observed for the first time in the $fp$ shell, the $\beta$-delayed $\gamma$-proton emission. Here we provide more detail of the data analysis, not discussed in Ref. \cite{Orrigo2014}, that is important for the proper determination of the $\beta$-decay strengths. We focus on the determination and subsequent analysis of the charged-particle spectrum measured by the Double-Sided Silicon Strip Detector (DSSSD).
	
\section{The experiment}
\begin{figure}[!h]
  	\centering
   	\includegraphics[width=0.55\columnwidth,trim={0 0 0 3.9cm},clip]{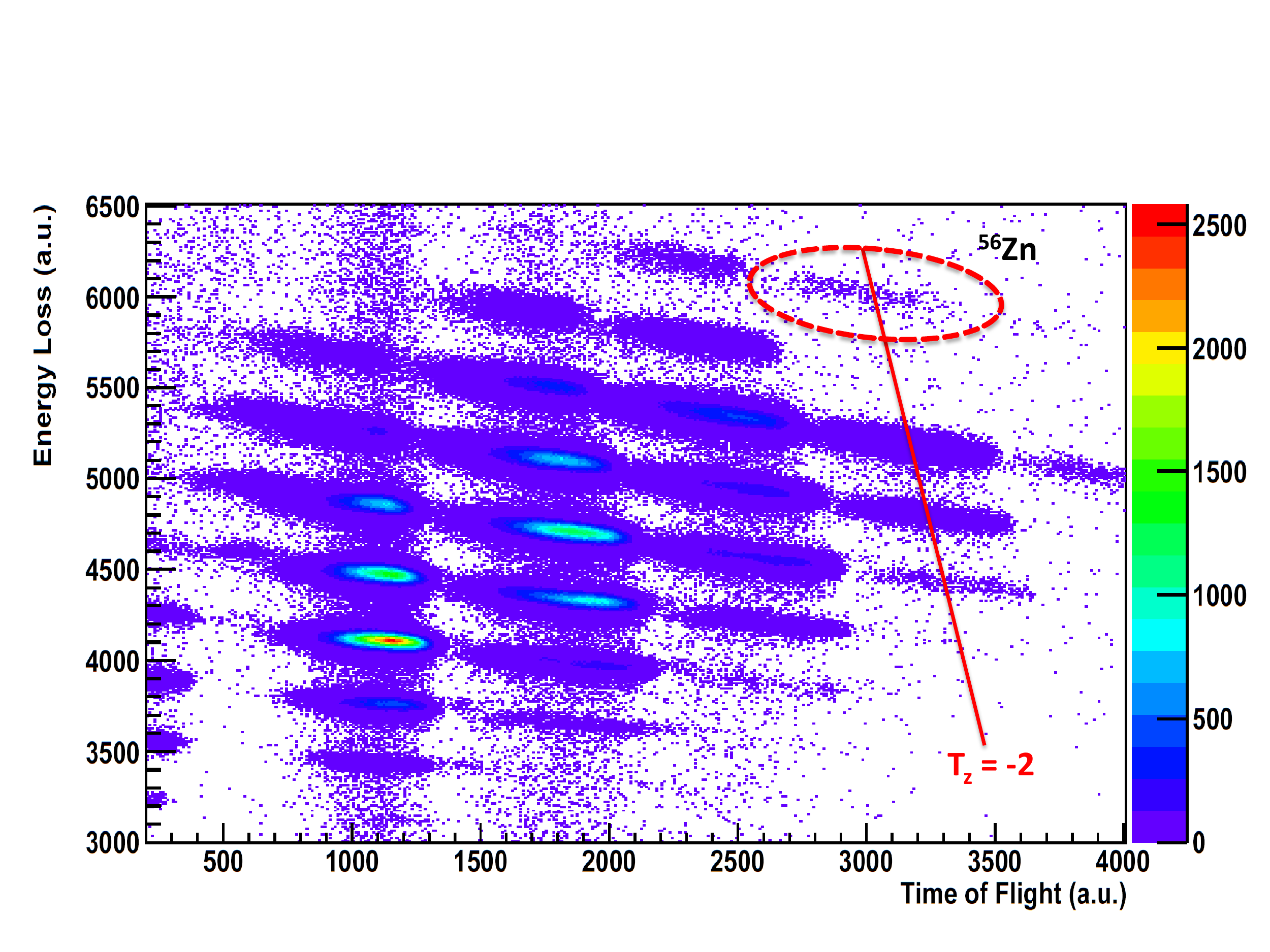}
  	\caption{$\Delta E$ versus ToF identification plot showing the position of the $^{56}$Zn implants.}
  	\label{IDplot}
	\end{figure}

The $\beta$-decay experiment was performed at GANIL using a $^{58}$Ni$^{26+}$ primary beam of 3.7 e$\mu$A, accelerated to 74.5 MeV/nucleon and fragmented on a 200 $\mu$m thick $^{nat}$Ni target. The fragments were selected by the LISE3 separator and implanted into a DSSSD (300 $\mu$m thick), surrounded by four EXOGAM Ge clovers for $\gamma$ detection. The DSSSD had 16 X and 16 Y strips, defining 256 pixels. The DSSSD was used to detect both the implanted fragments and subsequent charged-particle decays, by employing two parallel electronic chains of different gain. An implantation event was defined by simultaneous signals in both a silicon $\Delta E$ detector located upstream and the DSSSD. The implanted ions were identified by combining the energy loss signal in the $\Delta E$ detector and the Time-of-Flight (ToF) defined as the time difference between the cyclotron radio-frequency and the $\Delta E$ signal (see Fig. \ref{IDplot}). Decay events were defined as giving a signal above threshold (50-90 keV) in the DSSSD and no coincident $\Delta E$ signal.

\vspace{-5mm}

\section{The DSSSD charged-particle spectrum}
\begin{figure}[!ht]
	\begin{minipage}{0.5\columnwidth}
	  \centering
	  \includegraphics[width=1\columnwidth]{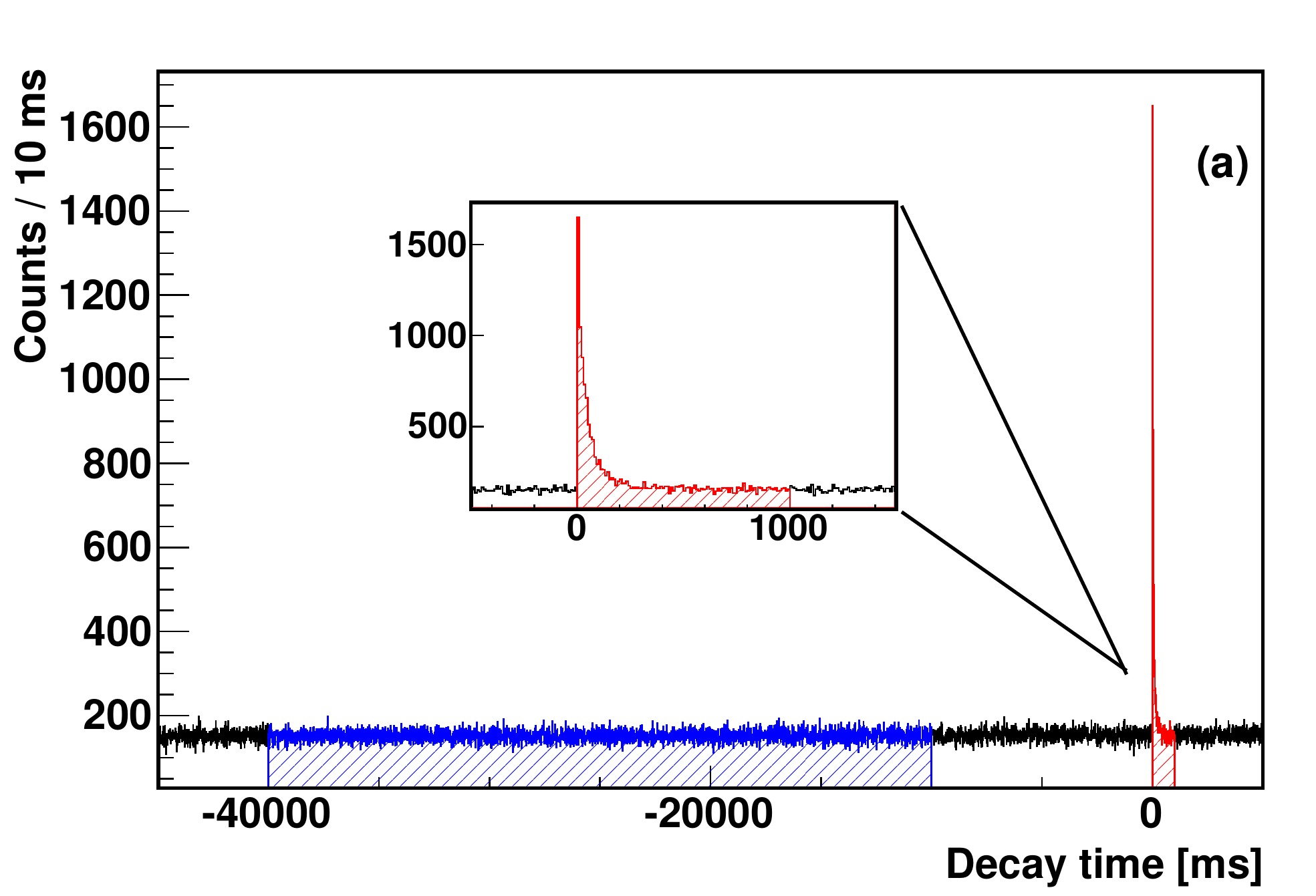}
  \end{minipage}
	\begin{minipage}{0.5\columnwidth}
	  \centering
	  \includegraphics[width=1\columnwidth]{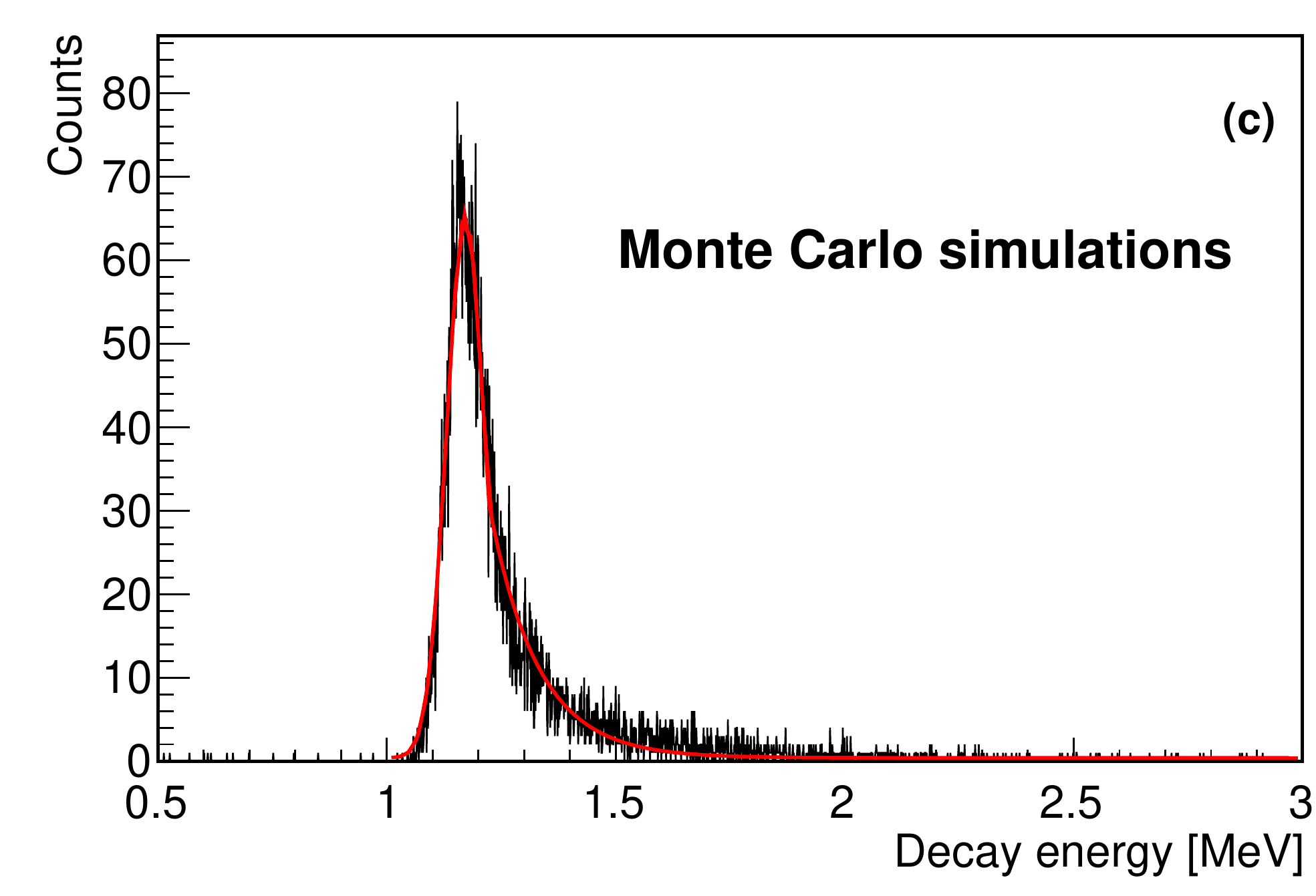}
	\end{minipage}
	\begin{minipage}{1\columnwidth}
		\centering
	  \includegraphics[width=0.6\columnwidth]{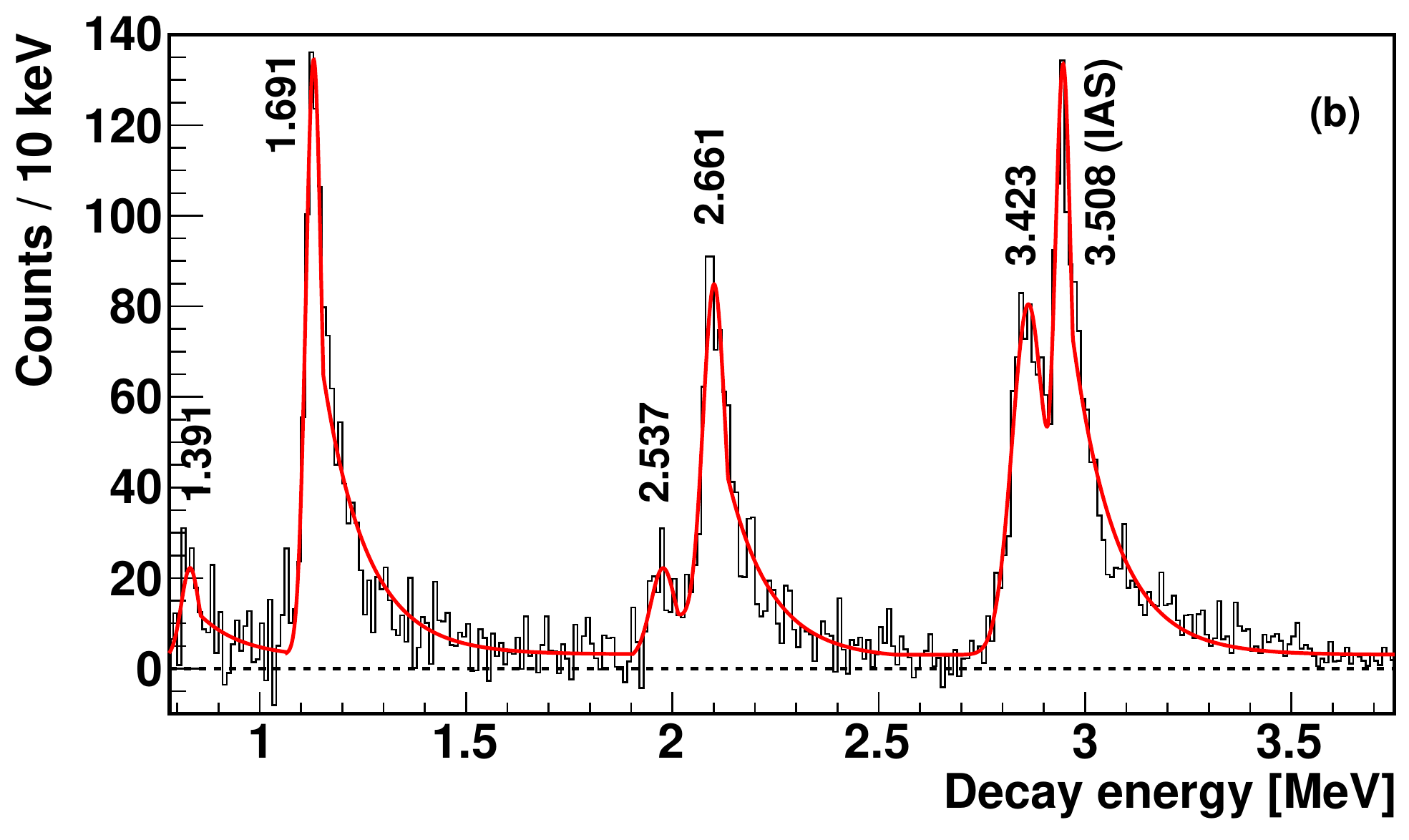}
	\end{minipage}
	\caption{a) Time correlations between each decay and all the $^{56}$Zn implants. The 0 to 1 s and -40 to -10 s gates are indicated by the red and blue regions, respectively. b) DSSSD charged-particle spectrum for decay events correlated with $^{56}$Zn implants. The peaks are labeled according to the corresponding excitation energies in $^{56}$Cu. c) Monte Carlo simulations of the level at $E_{p}$ = 1.1 MeV. }
	\label{fig2}
\end{figure}

The $\it{correlation~time}$ is defined as the time difference between a decay event in a given pixel of the DSSSD and any implantation signal that occurred before and after it in the same pixel and also satisfied the conditions required to identify the nuclear species. This procedure ensured that all the true correlations were taken into account. However many random correlations were also included producing, as expected, a large constant background. The correlation-time spectrum for $^{56}$Zn including all the decays ($\beta$s and protons) is shown in Fig. \ref{fig2}a.

The DSSSD charged-particle spectrum was formed as it follows. First a spectrum containing both true and random correlations was formed by setting a gate from 0 to 1 s on the time correlation (Fig. \ref{fig2}a). A background (bg) spectrum formed only of randoms was created by selecting time correlations from -40 to -10 s. The two spectra were then normalized to the time bin used, and the DSSSD spectrum in Fig. \ref{fig2}b was formed by subtracting them. This procedure is similar to that of Ref. \cite{Dossat2007}, with the difference that there the gate for the bg spectrum was chosen from 1 to 2 s. Using a larger time gate (30 s) we have increased the statistics of the bg spectrum, which has the advantage of reducing the fluctuations arising from the subtraction of the two spectra. This is important in a case such as $^{56}$Zn where the number of counts is low. Moreover, the choice of an interval on the left of the time peak ensures that only randoms are included in the bg spectrum.

The DSSSD spectrum was calibrated using an $\alpha$-particle source and the peaks of known energy from the decay of $^{53}$Ni \cite{Dossat2007}. Most of the strength in Fig. \ref{fig2}b is interpreted as $\beta$-delayed proton emission. The resolution is limited by the summing with the coincident $\beta$ particles, which also affects the lineshape of the peak. Monte Carlo simulations were performed for a Silicon DSSSD strip, using the 4.9.6 Geant4 code. The radioactive sources were located in an extended area in the middle of the detector, with an implantation profile obtained from LISE calculations. Protons of a given energy $E_{p}$ were emitted at the same time as $\beta$ particles following a distribution determined by the Fermi function ($\beta$-decay event generator), with end-point energy corresponding to $Q_{\beta}-E_{p}-S_{p}$ (where $S_{p}$ is the proton separation energy in the daughter nucleus). At the event generation level a widening of 70 keV FWHM was imposed, corresponding to the DSSSD experimental resolution. As an example, the result of the simulations for a single level at $E_{X}$ = 1.7 MeV ($E_{p}$ = 1.1 MeV) is shown in Fig. \ref{fig2}c. To a good approximation the lineshape is given by a Gaussian plus an exponential high-energy tail. This result is an additional confirmation of the procedure widely used in Ref \cite{Dossat2007}. The lineshape obtained from the simulations, which was also checked with the well isolated $^{57}$Zn peak at $E_{p}$ = 4.6 MeV, was then used to fit the experimental spectrum (Fig. \ref{fig2}b). The shape was kept fixed (i.e., the Gaussian-exponential joining and the slope of the exponential), while the parameters of the Gaussian (height, mean, sigma) were fitted to the experimental proton peaks. This procedure allows one to determine the intensities of the proton peaks properly and hence obtain the $\beta$-decay strengths.

This work was supported by the Spanish MICINN grants FPA2008-06419-C02-01, FPA2011-24553; CPAN Consolider-Ingenio 2010 Programme CSD2007-00042; MEXT, Japan 18540270 and 22540310; Japan-Spain coll. program of JSPS and CSIC; Istanbul University Scientific Research Projects, Num. 5808; UK Science and Technology Facilities Council (STFC) Grant No. ST/F012012/1; Region of Aquitaine. R.B.C. acknowledges support from the Alexander von Humboldt foundation and the Max-Planck-Partner Group. We acknowledge the EXOGAM collaboration for the use of their clover detectors.

\end{document}